\documentclass[pre,showpacs,a4paper]{revtex4}
\usepackage{multibox,bm}
\def\p{\partial}
\def\a{\alpha}
\def\d{\delta}
\def\e{\varepsilon}

\def\l{\lambda}
\def\r{\rho}
\def\g{\gamma}
\def\o{\omega}

\def\z{\zeta}
\def\s{\sigma}
\def\ra{\rightarrow}

\def\O{\Omega}
\def\bO{{\bf \Omega}}

\def\T{{\tilde T}}
\begin{document}
\title{Anisotropy in Homogeneous Rotating Turbulence}
\author{Jos\'e Gaite} 
\affiliation{Instituto de Matem{\'a}ticas y F{\'\i}sica Fundamental,
CSIC, Serrano 123, 28006 Madrid, Spain}

\date{October 22, 2003}

\begin{abstract}
The effective stress tensor of a homogeneous turbulent rotating fluid
is anisotropic. This leads us to consider the most general
axisymmetric four-rank ``viscosity tensor'' for a Newtonian fluid and
the new terms in the turbulent effective force on large scales that
arise from it, in addition to the microscopic viscous force.  Some of
these terms involve couplings to vorticity and others are angular
momentum non conserving (in the rotating frame).  Furthermore, we
explore the constraints on the response function and the two-point
velocity correlation due to axisymmetry.  Finally, we compare our
viscosity tensor with other four-rank tensors defined in current
approaches to non-rotating anisotropic turbulence.
\end{abstract}


\maketitle

\section{Introduction}
\label{intro}

The properties and applications of rotating fluids constitute an
important area of fluid mechanics \cite{Greenspan}.  In particular,
the anisotropy consequent to the rotation has been a very important
subject. For example, we have the classical Proudman-Taylor theorem,
which says that, in the limit of fast rotation, the flow is so
strongly anisotropic that actually becomes two-dimensional.
Turbulence in the presence of uniform rotation, which is called
rotating turbulence, is an example of anisotropic turbulence and an
area of active research \cite{Greenspan,Cambon1,Jacquin,Cambon2}.

From a theoretical point of view regarding symmetry, the classical
theory of fully developed turbulence assumes the maximal possible
symmetry, namely, symmetry under translations and rotations, so it
applies to ideal {\em homogeneous and isotropic} turbulence.  However,
in various situations such a high symmetry is not realistic and one
must consider less symmetric turbulent states. The next most symmetric
state is still homogeneous but the isotropy reduces to axisymmetry,
that is, the full rotation group reduces to the group of rotations
around a particular axis.  The archetype of homogeneous turbulence
with axisymmetry is rotating turbulence (naturally, the symmetry axis
is the rotation axis).

It was shown in Ref.\ \cite{us} that perturbation theory for the
randomly forced rotating Navier-Stokes equation generates anisotropic
effective forces, in particular, the {\em nondissipative} force $\bO
\times \nabla^2 {\bm u}$.  This suggests that one should find the
complete set of allowed force terms.  In this regard, it is useful to
define the effective viscosity, which is a tensorial function of $\bO$
(reproducing the known perturbative results for isotropic turbulence
as $\Omega \ra 0$).  Beyond perturbation theory (or any method of
solution), our intention here is to study form first principles the
consequences of axisymmetry in rotating turbulence.

The possibility of anisotropy in the velocity correlation functions
has been considered before in non-rotating
fluids~\cite{Chandra,DuFri,Frisch,Slov,Arad,jap,Proca}.  So, in these
references, the anisotropy was attributed to other causes: existence
of a mean flow or anisotropic forcing. In fact, in a {\em homogeneous}
fluid the existence of mean flow effects, that is, the dependence of
properties of the flow on its mean velocity, would contradict Galilean
invariance. The flow can only depend on global kinematical features
that involve accelerations, such as in a uniformly rotating fluid.  A
homogeneous but anisotropic forcing will induce anisotropy in the
velocity field (the axial case is studied in Ref.\ \cite{Slov}); but
the physical origin of this anisotropic forcing and, therefore, the
extent of the scale range affected by it are not clear.  We think that
rotating turbulence is a more natural example of anisotropic
turbulence and with more physical applications.  Moreover, this type
of anisotropic turbulence has distinctive features (as was pointed out
in Ref.\ \cite{us}) because $\bO$ is an axial vector.  Indeed, the
force $\bO \times \nabla^2 {\bm u}$ or other terms of the same type
would not be allowed if isotropy were broken by a polar vector, as in
Ref.\ \cite{Slov}.

We remark that the characterization of the effective (or eddy)
viscosity as a four-rank tensor has already appeared in the
literature.  For example, in Ref.\ \cite{DuFri} the authors show that
a multi-scale method applied to the Navier-Stokes equation linearized
with respect to a weak large-scale flow precisely produces an
effective viscosity tensor if the basic fluctuating flow is not
isotropic.  However, as commented above, to determine the form of this
tensor, one needs an explicit mechanism that breaks isotropy and
preserves homogeneity. Otherwise, the basic assumptions and, in
particular, axisymmetry, are not justified.

On the other hand, since $\bO$ is an axial vector, the effective
viscosity tensor in rotating turbulence has distinctive features: for
example, it has a pair-antisymmetric piece (which generates the above
mentioned force) \cite{us}.  In addition, it will be shown here that a
general treatment of the efective viscosity tensor in rotating
turbulence requires new terms that couple to the vorticity or that are
angular momentum non conserving (in the rotating frame) and,
therefore, are forbidden in anisotropic non-rotating turbulence.

We shall first review the fluid equations in the rotating frame and
the conditions for turbulence; we emphasize the transition from small
scale isotropic turbulence to large scale anisotropic turbulence.
Next, we introduce the viscosity in the standard manner \cite{LL} but
{\em without} recourse to isotropy, which is replaced by only
axisymmetry. All the components of the resulting four-rank tensor are
determined with group theory arguments \cite{Ham}.  From this tensor
we obtain the additional anisotropic force terms.  Once seen the
axisymmetry constraints on the viscosity tensor, we impose axisymmetry
on the response function or (two-point) velocity correlations.  In
particular, the large-scale response function is related with the
viscosity tensor.  Finally, we try to connect the viscosity tensor
with other four-rank tensors introduced in some current approaches to
non-rotating anisotropic turbulence.

\section{Equations of motion in a rotating frame and turbulence}
\label{sec:equations}

The hydrodynamical equations for a fluid with density field $\rho({\bm
x},t)$, velocity field ${\bm u}({\bm x},t)$, pressure $P({\bm x},t)$
in a frame rotating with constant angular velocity ${\bf \Omega}$ are
\begin{eqnarray}
\frac{\partial \rho}{\partial t} + \nabla \cdot(\rho {\bm u})
&=& 0
\; ,
\label{density1}
\\
\frac{\partial {\bm u}}{\partial t} + ({\bm u }\cdot \nabla)
{\bm u} &=& -\frac{1}{\rho} \nabla P
-2{\bf \Omega \times u}
- {\bO \times (\bO \times {\bm x})}
+ {\bm  f}
\; ,
\label{moment1}
\end{eqnarray}
where ${\bm f}$ accounts for an additional acceleration due to
friction (which vanishes if ${\partial_i{\bm u} = 0}$) and a
homogeneous and isotropic external forcing, usually  
random (or periodic as in Ref.~\cite{DuFri}), 
which serves for keeping the total kinetic energy constant.

We assume that the fluid is incompressible, with constant density, so
the continuity equation becomes $\nabla \cdot {\bm u}=0$.  So if we
define $p = P/\rho$ every reference to the density disappears, and we
have two equations for the two unknowns ${\bm u}$ and $p$.  To solve
for $p$, it is useful to separate Eq.~(\ref{moment1}) into independent
longitudinal and transverse equations. Since ${\bm u}= {\bm u}_L +
{\bm u}_T$ (such that $\nabla \times{\bm u}_L = \nabla \cdot {\bm
u}_T=0$) and ${\bm u}_L$ identically vanishes, the longitudinal
equation becomes just a constraint relating $p$ with spatial
derivatives of ${\bm u}$, namely,
\begin{equation}
p = \frac{1}{2} ({\bf \Omega} \times {\bm x})^2
-\frac{1}{\nabla^2}
\left[
\partial_i \left( u_j \partial_j u_i \right)
+ 2\epsilon_{ijk} \Omega_j \partial_i u_k
\right]
\; .
\label{p}
\end{equation}
Solving for $p$, the equation for ${\bm u}= {\bm u}_T$ is
\begin{eqnarray}
\frac{\partial \bm u}{\partial t} &+&
{\cal P}
[({\bm u \cdot  \nabla}) {\bm u}] = 
- {\cal P}(2{\bO \times {\bm u}}) + {\bm f}
\; ,
\label{u-field2}
\end{eqnarray}
where the projection operator ${\cal P}$ onto transverse (or solenoidal) 
fields is given by
\begin{eqnarray}
{\cal P} =  {\bf 1} - {\nabla} \frac{1}{\nabla^2}{\nabla} 
\cdot  \; . \label{project}
\end{eqnarray}

In Eq.~(\ref{u-field2}), if ${\bm u}$ is tranverse so
is ${\bm f}$ and vice versa. We call Eq.\ (\ref{u-field2}) the
transverse rotating fluid equation. If we substitute for ${\bm f}$ an
isotropic viscous force, it becomes the transverse rotating
Navier-Stokes equation.  Note that the transverse rotating fluid
equation (\ref{u-field2}) is {\em translation invariant} (assuming
that ${\bm f}$ is homogeneous), in contrast with Eq.\ (\ref{moment1}).
Therefore, its solutions are homogeneous velocity fields and,
furthermore, one can make use of the Fourier transform.

\subsection{Homogeneous rotating turbulence}

The homogeneous rotating turbulent state is defined by a velocity
field with large fluctuations but such that the mean velocity is
negligible in the rotating frame.  Let us see how to characterize this
state in terms of nondimensional parameters and how it is related with
the homogeneous and isotropic turbulent state.

Since we have the rotation velocity as additional parameter, we can
define two nondimensional parameters, namely, the Reynolds and Rossby
numbers.  While the Reynolds number $Ro = UL/\nu$ measures the
relative importance of the nonlinear and viscosity terms in the
Navier-Stokes equation, the Rossby number $Ro = U/(L \O)$ measures the
relative importance of the nonlinear and Coriolis terms in the
rotating Navier-Stokes equation ($U$ is a reference velocity or the
variation of the velocity over the length $L$ that characterizes the
system size).  In principle, $Ro \gg 1$ indicates that rotation
effects are negligible and, viceversa, $Ro \ll 1$ indicates that they
are dominant.  However, the latter condition, namely, dominance of
rotation effects over nonlinear and viscous (and even dynamic) effects
leads to the situation in which only the linear Coriolis force is
relevant, giving rise to extreme two-dimensionalization of the flow
(as in the Proudman-Taylor theorem) but without turbulence.  It is
necessary that the {\em two} numbers $Ro$ and $Re$ play a role in
specifying the regime of interest, that is to say, the regime with
rotation effects ($Ro \gg 1$) {\em and} turbulence ($Re \gg 1$).  Or
one may introduce the Ekman number $Ek = \nu/(\Omega L^2)$ (in
addition to $Ro$), which is the ratio of the Rossby number to the
Reynolds number and measures the relative importance of the viscosity
and Coriolis terms \cite{Greenspan}.  Then one must demand $Ek \ll Ro$
in addition to $Ro \ll 1$.

To clarify the preceding condition, let us consider relevant length
scales. First, let us recall the role of the dissipation scale.  In
ordinary homogenous and isotropic turbulence, K41 theory \cite{Frisch}
makes the dissipation rate per unit mass $\e$ the basic quantity and
introduces the dissipation scale $\l=(\nu^3/\e)^{1/4}$. Using $\O$
instead of $\nu$, we can form with $\e$ the length scale $\ell =
(\e/\O^3)^{1/2}$.  If we begin with small $\O$ (for fixed $\e$) such
that $\ell \gg L$, rotation effects must be negligible all over the
fluid system of characteristic length $L$.  Therefore, the precise
condition for neglecting $\O$ is $\O \ll (\e/L^2)^{1/3}$ (equivalent
to $\ell \gg L$).  Given that $(L/\l)^{4/3} = Re \gg 1$, the parameter
$(\e/L^2)^{1/3}/\O = \nu/(\Omega L^2) (L/\l)^{4/3} = Ek\,Re = Ro$, so
the condition for neglecting $\O$ is just $Ro \gg 1$.  As $\O$ grows
and, therefore, $\ell$ diminishes such that $\ell < L$, rotation
effects become appreciable. We then have one scale range with rotating
turbulence, namely, between $\ell$ and $L$, and another with isotropic
turbulence, namely, between $\l$ and $\ell$.  The latter range holds
as long as $\l < \ell$, that is, $\O < \sqrt{\e/\nu}$. As $\ell$
becomes smaller than $\l$, the rotation effects dominate over the
nonlinear effects and the flow becomes strongly two-dimensional.

The interesting values of $\O$ are such that there are the two scale
ranges, respectively, with isotropic turbulence on small scales and
anisotropic turbulence on larger scales. Of course, this happens when
$\l \ll \ell \ll L$. Then the viscosity or correlation functions on
scales between $\l$ and $\ell$ are essentially isotropic whereas the
effective viscosity or correlation functions on scales between $\ell$
and $L$ are axisymmetric.

\section{The axisymmetric effective viscosity tensor}
\label{sec:effective-tensor}

To introduce the viscosity tensor, it is convenient to follow the
general reasoning \cite{LL} which starts by writing the fluid equation
in local conservative form, as
\begin{equation}\label{funda}
\rho \frac{\p u_i}{\p t} =  \frac{\partial \Pi_{ij}}{
\partial x_j}, \quad \Pi_{ij} = -\r u_i u_j + T_{ij} \;,
\end{equation}
and finds the deviatoric part of
the stress tensor
\begin{equation}\label{decompo}
T_{ij} = -P\, \delta_{ij} + \sigma_{ij}
\end{equation}
due to internal relative motion (viscosity) from general
principles. The first principle is that the velocity gradient is
small, which allows one to consider only {\em first derivatives} of
the velocity. Next, the viscous stress tensor $\sigma_{ij}$ is taken
proportional to the velocity gradient and, furthermore, its
antisymmetric components (vorticity) are excluded, so that the stress
is proportional to the rate of strain $u_{mn} = \p_{(m} u_{n)} =
(\partial u_n/\partial x_m + \partial u_m/\partial x_n)/2$ (this
characterizes {\em Newtonian fluids})
\footnote{We use the common notation for symmetrization or 
antisymmetrization over two indices: $A_{(ij)} = (A_{ij} + A_{ji})/2$
and $A_{[ij]} = (A_{ij} - A_{ji})/2$.}.
The following crucial assumption is {\em isotropy},
which leads to the existence of only two proportionality constants 
(shear and bulk viscosities). As we cannot make this assumption here, 
we are left with just the proportionality relation  
\begin{equation}
\s_{ij} = \eta_{ijmn} \,u_{mn}
\, ,
\label{newton}
\end{equation}
like in the analogous relation in the theory of elasticity that
expresses that the stress is proportional to the strain~\cite{LL2}.
Therefore, the symmetry properties of the tensor $\eta_{ijmn}$, which
we call the ``viscosity tensor'', are similar to the ones of the
elastic modulus tensor, namely, symmetry under exchange of indices
within the first and second pairs of indices and, in addition,
symmetry under exchange of the first and second pairs of indices (pair
symmetry).  However, we shall further allow for {\em pair
antisymmetry}; namely, we write $\eta_{ijmn}$ as a sum of a
pair-symmetric (S) and a pair-antisymmetric (A) part \cite{us}:
\begin{eqnarray}
\eta_{ijmn}
&=&
\frac{1}{2}(\eta_{ijmn} +\eta_{mnij} )
+
\frac{1}{2}(\eta_{ijmn} - \eta_{mnij} )
\equiv
\eta^{S}_{ijmn} + \eta^{A}_{ijmn}
\; .
\label{viscosity}
\end{eqnarray}
So, generically, the ``viscosity tensor'' has $36$ independent
components, of which $21$ belong to the pair-symmetric part
$\eta^{S}_{ijmn}$ and $15$ belong to the pair-antisymmetric part
$\eta^{A}_{ijmn}$.

The axial symmetry of the equations of motion reduces the number of
independent components of both $\eta^S_{ijmn}$ and
$\eta^A_{ijmn}$. The 21 components of the generic pair-symmetric tensor
can be divided into two sets with 15 and 6 components, respectively,
the former corresponding to the totally symmetric tensor. The
respective components are constructed in the appendix as linear
representations and called $_{15}S$ and $_{6}S$. Further imposing
axisymmetry, the pair-symmetric tensor can be constructed from
$\Omega_i$ and $\delta_{ij}$ as
\begin{eqnarray}
\eta_{ijmn}^S
&=&
a_1 (\delta_{ij}\delta_{mn} + 
\delta_{im}\delta_{jn} + \delta_{in}\delta_{jm})
\nonumber
\\
&+&
a_2 (\Omega_{i} \Omega_j \delta_{mn}
+\Omega_{m} \Omega_n \delta_{ij}
+\Omega_{i} \Omega_m \delta_{jn}
+\Omega_{j} \Omega_m \delta_{in}
+\Omega_{i} \Omega_n \delta_{jm}
+\Omega_{j} \Omega_n \delta_{im})
\nonumber
\\
&+&
a_3 \Omega_{i} \Omega_{j}\Omega_{m}\Omega_{n}
\nonumber
\\
&+&
a_4 \delta_{ij}\delta_{mn}
\nonumber
\\
&+&
a_5 (\Omega_{i} \Omega_j \delta_{mn}
+\Omega_{m} \Omega_n \delta_{ij})
\; .
\label{nus}
\end{eqnarray}
There are five independent components, to which we attach scalars
$a_1, \ldots, a_5$ (which can depend on $\Omega^2$). In comparison
with the form given in Ref.\ \cite{us}, this expression has been
arranged so that the three first tensors (with coefficients $a_1, a_2,
a_3$) are totally symmetric in their indices.

The generic pair-antisymmetric tensor has 15 components,
constructed in the appendix as the linear representation $_{15}S'$.
The pair-antisymmetric tensor with axisymmetry needs, in addition to
$\Omega_i$ and $\delta_{ij}$, the totally antisymmetric tensor
$\epsilon_{ijk}$ and is
\begin{eqnarray}
\eta_{ijmn}^A
&=&
b_1 \Omega_q
(\epsilon_{qim}\delta_{jn} +
\epsilon_{qin}\delta_{jm} +
\epsilon_{qjm}\delta_{in} +
\epsilon_{qjn}\delta_{im} )
\nonumber
\\
&+&
b_2 \Omega_q
(\epsilon_{qim}\Omega_{j}  \Omega_{n}
+\epsilon_{qin} \Omega_{j}  \Omega_{m}  +
\epsilon_{qjm}\Omega_{i}  \Omega_{n}  +
\epsilon_{qjn}\Omega_{i}  \Omega_{m}  )
\nonumber
\\
&+&
b_3 (\Omega_{i} \Omega_j \delta_{mn}
-\Omega_{m} \Omega_n \delta_{ij})
\label{nua}
\; .
\end{eqnarray}

We observe that the axisymmmetry has reduced the number of independent
components from $21$ to $5$ for the pair-symmetric part and from $15$
to $3$ for the pair-antisymmetric part. This reduction can be
explained by considering the reduction of linear tensor
representations under rotations (see appendix and Ref.\ \cite{Arad}).
The reduction under rotations is performed by extracting traces, which
are rotation invariant {\em but not} linear invariant. There exists a
canonical procedure for doing this trace extraction \cite{Ham} but we
can clarify the procedure by noting that the properties of the
expressions in Eq.\ (\ref{nus}) or Eq.\ (\ref{nua}) under rotations
are determined only by the vector $\O_i$ ($\delta_{ij}$ and
$\epsilon_{ijk}$ are rotation invariant). For example, the terms with
coefficients $a_1$ and $a_4$ clearly correspond to scalars ($J=0$),
the only terms allowed by isotropy.  Furthermore, an expression with
the tensor product of $n$ $\bO$'s corresponds to the representation
$J=n$, usually, with an admixture of lower $J$ representations.
Therefore, each coefficient corresponds to a definite $J$
representation, but, in order to obtain the correct tensorial
expression of each representation, we need to remove the lower $J$
representations by extracting traces. This induces a linear
redefinition of the coefficients within each linear representation.

Finally, we remark that the terms in Eq.\ (\ref{nua}) with
coefficients $b_1$ and $b_2$ would not be allowed if isotropy were
broken by a polar vector, because the respective terms would be odd
under parity (in general, the parity of the representation $J$
associated to a polar vector is $(-)^J$).

\subsection{Traceless components and incompressibility constraint}

The preceding tensorial expressions for the viscosity have a part that
couples to the velocity divergence $u_{ii}$. Moreover, they give rise
to an isotropic part of the viscous stress tensor $\s_{ij}$ (that is,
proportional to $\d_{ij}$). Therefore, the viscosity tensors must be
further decomposed into traceless and trace parts. For incompressible
flow we only need the traceless components such that $\eta_{ijkk} =
\eta_{kkmn} = 0$. They can be extracted by subtracting traces from
either $\eta_{ijmn}^S$ or $\eta_{ijmn}^A$. In the general case, that
is, with no axial (or any other) symmetry, those tracelessness
conditions remove $6 + 6 -1 = 11$ components (the condition
$\eta_{kkll} = 0$ appears twice), leaving 25 components.  To be more
precise, the conditions $\eta^S_{ijkk} = 0$ remove 6 components of
$\eta_{ijmn}^S$ (with $J=2,0$, corresponding to $_{6}S$), and the
conditions $\eta^A_{ijkk} = 0$ remove the 5 components of
$\eta_{ijmn}^A$ corresponding to $J=2$.

Indeed, a straightforward calculation yields:
\begin{eqnarray}
\eta_{ijmn}^S - \frac{1}{3}\eta_{ijkk}^S \delta_{mn}
 - \frac{1}{3}\eta_{kkmn}^S \delta_{ij}
+ \frac{1}{9}\eta_{kkll}^S \delta_{ij}\delta_{mn}
= a_1 (\delta_{im}\delta_{jn} + \delta_{in}\delta_{jm}
- \frac{2}{3} \delta_{ij}\delta_{mn}) +
\nonumber
\\
a_2 [\Omega_{i} \Omega_m \delta_{jn}
+\Omega_{j} \Omega_m \delta_{in}
+\Omega_{i} \Omega_n \delta_{jm}
+\Omega_{j} \Omega_n \delta_{im}
- \frac{4}{3} (\Omega_{i} \Omega_j \delta_{mn}
+\Omega_{m} \Omega_n \delta_{ij})
+ \frac{4}{9} \Omega^2 \delta_{ij}\delta_{mn}
]
\nonumber
\\
{}+ a_3 [\Omega_{i} \Omega_{j}\Omega_{m}\Omega_{n}
- \frac{1}{3} \Omega^2 (\Omega_{i} \Omega_j \delta_{mn}
+\Omega_{m} \Omega_n \delta_{ij})
+ \frac{1}{9} \Omega^4 \delta_{ij}\delta_{mn}
],
\\
\eta_{ijmn}^A - \frac{1}{3}\eta_{ijkk}^A \delta_{mn}
 - \frac{1}{3}\eta_{kkmn}^A \delta_{ij}
= b_1 \Omega_q
(\epsilon_{qim}\delta_{jn} +
\epsilon_{qin}\delta_{jm} +
\epsilon_{qjm}\delta_{in} +
\epsilon_{qjn}\delta_{im} )
\nonumber
\\
{}+ b_2 \Omega_q
(\epsilon_{qim}\Omega_{j}  \Omega_{n}
+\epsilon_{qin} \Omega_{j}  \Omega_{m}  +
\epsilon_{qjm}\Omega_{i}  \Omega_{n}  +
\epsilon_{qjn}\Omega_{i}  \Omega_{m}  )
.
\label{tnu}
\end{eqnarray}
The number of coefficients has been reduced to three for
$\eta_{ijmn}^S$, corresponding to the $J=4,2,0$ representations, and
to two for $\eta_{ijmn}^A$, corresponding to $J=3,1$.  It is natural
that they together constitute the Clebsh-Gordan decomposition of the
tensor product of two $J=2$ representations (with dimension $5 \times
5 = 25$) \cite{Ham}.

There is another set of tracelessness conditions, namely, $\eta_{ijmj}
= 0$, but there is no physical reason to impose them.  However, note
that the six conditions $\eta^S_{ijmj} = 0$ remove the $J=2,0$
representations, just leaving $J=4$, while the three conditions
$\eta^A_{ijmj} = 0$ remove the $J=1$ representation, just leaving
$J=3$. Therefore, this last set of tracelessness conditions would
select the highest $J$ representations, corresponding to the
coefficients $a_3$ and $b_2$.

\subsection{Viscosity tensors with antisymmetric pairs}

Two crucial assumptions in the reasoning at the beginnning of Sect.\
\ref{sec:effective-tensor} are that the viscous stress tensor is
symmetric and that it does not depend on the vorticity (the vorticity
tensor is $\o_{ij} = \p_{[i} u_{j]} = (\p_i u_j - \p_j u_i)/2$).  They
lead to a viscosity tensor with symmetry under exchange of indices
within the first and second pairs of indices (symmetry by
pairs). Those two assumptions are commonly accepted since they are
based on basic physical principles: on the one hand, the stress tensor
can always be chosen symmetric because of angular momentum
conservation; on the other hand, a uniform rotation (leading to a
constant vorticity) cannot induce stresses, so a dependence of the
stress tensor on vorticity is forbidden. However, both principles,
namely, angular momentum conservation and absence of stresses in
uniformly rotating fluid, fail in a rotating frame.  Therefore, we are
allowed to consider viscosity tensors with antisymmetric pairs of
indices.  We have three types: (i) tensors $\chi_{ijmn}$ with the
first pair symmetric and the second antisymmetric, which account for
an angular momentum conserving coupling to vorticity, (ii) the
symmetric type $\xi_{ijmn}$, that is, tensors with the first pair
antisymmetric and the second symmetric, which account for an angular
momentum non-conserving coupling to strain rate, and (iii) tensors
$\zeta_{ijmn}$ with both pairs antisymmetric, which account for an
angular momentum non-conserving coupling to vorticity.

The most general tensor with the first pair symmetric and the second
antisymmetric has 18 components (see the appendix).  Its axisymmetric
form is
\begin{eqnarray}
\chi_{ijmn} = (c_1 \delta_{ij} + c_2 \O_i \O_j) \epsilon_{lmn}\O_l +
\nonumber\\
c_3(\O_i \O_m \d_{jn} + \O_j \O_m \d_{in} - \O_i \O_n \d_{jm} 
- \O_j \O_n \d_{im})
+ c_4 (\epsilon_{imn}\O_j + \epsilon_{jmn}\O_i)
\,.
\label{chi}
\end{eqnarray}
The constants $c_1,c_3,c_2$ correspond to $J=1,2,3$, respectively,
forming the linear representation $_{15}S\!A$, whereas $c_4$
corresponds to $J=1$ and $_{3}S\!A$.  Imposing that the tensor be
traceless in its first two indices, that is, $\chi_{iimn}=0$, relates
the coefficients $c_1$ and $c_2$ (the tensor is automatically
traceless in the second pair of indices).  Therefore, the traceless
tensor contains the $J=1,2,3$ representations, corresponding to the
Clebsh-Gordan decomposition of the tensor product of the $J=2$ and
$J=1$ representations.

There is an analogous axisymmetric structure for the symmetric type
$\xi_{ijmn}$, involving $_{15}AS$ and $_{3}AS$, and with coefficients
$c'_1, \ldots, c'_4$.

Finally, the tensor $\zeta_{ijmn}$ with both pairs antisymmetric has 9
components, which the axisymmetry reduces to
\begin{eqnarray}
\zeta_{ijmn} = d_1 (\delta_{im} \delta_{jn} - \delta_{in} \delta_{jm}) +
\nonumber\\
d_2(\O_i \O_m \d_{jn} - \O_j \O_m \d_{in} - \O_i \O_n \d_{jm} 
+ \O_j \O_n \d_{im})
+ d_3 (\epsilon_{imn}\O_j - \epsilon_{jmn}\O_i)
\,.
\label{zeta}
\end{eqnarray}
The constants $d_1,d_2$ correspond to $J=0,2$, respectively, forming
the representation $_{6}A$ (which is pair symmetric), whereas $d_3$
corresponds to $J=1$ and $_{3}A$ (which is pair antisymmetric: even
though it may not seem obvious, $\epsilon_{imn}\O_j -
\epsilon_{jmn}\O_i = -\epsilon_{mij}\O_n + \epsilon_{nij}\O_m$).  The
tensor defined by Eq.\ (\ref{zeta}) is trivially traceless in both
pairs of indices and corresponds to the Clebsh-Gordan decomposition of
the tensor product of two $J=1$ representations.

\subsection{Effective forces associated with the viscosity tensor}

The total viscosity tensor $\tau = \eta + \chi + \xi + \z$
is defined by
\begin{equation}
\s_{ij} =  \tau_{ijmn} \, \p_m u_{n}\,.
\end{equation}
The force derived from this stress tensor is
\begin{equation}
f_i = \p\s_{ij}/\p x_j =  \tau_{ijmn} \,\p_{jm} u_{n}\,.
\label{force}
\end{equation}
The expression that results by substituting the full axisymmetric
expression of $\tau_{ijmn}$ is fairly complicated: suppressing
gradient terms, we obtain
\begin{eqnarray}\label{pseudo}
{\bm f} 
&=& (a_1 - d_1)\nabla^2{\bm u} -b_1 \, ({\bf\Omega}\times \nabla^2{\bm u}) 
- b_2 \,({\bf \Omega \cdot \nabla})^2 ({\bf \Omega} \times {\bm u}) 
- (b_2 + c'_2) \,({\bf \Omega \cdot \nabla}) ({\bf \Omega} 
\times {\bf \nabla}) ({\bf \Omega} \cdot {\bm u})\nonumber \\ 
&+& (b_2 + c_2)\,\bO\, ({\bf \Omega \cdot
  \nabla})( {\bf \Omega \cdot}{\bm \o}) + (c_4 + c'_4 + d_3)
  (\bO\cdot\nabla){\bm\o} 
  + (a_2 - c_3 + c'_3 - d_2)\bO\, \nabla^2 (\bO\cdot{\bm u}) 
  \nonumber \\ 
  &+& (a_2 + c_3 - c'_3 - d_2)(\bO\cdot\nabla)^2{\bm u}
  + a_3 \bO\,(\bO\cdot\nabla)^2(\bO\cdot{\bm u}).
\end{eqnarray}
Several remarks are in order.  Note that the fifth and sixth terms of
the force involve the vorticity ${\bm \o} = \nabla \times {\bm u}$ and
are proportional to an odd power of $\bO$. The terms preceding them
are also proportional to an odd power of $\bO$, except the first one,
which is isotropic.  The remainig three anisotropic terms, which
neither involve the vorticity nor any vector product, are equivalent
to the anisotropic force written in Ref.\ \cite{Slov}. If we had
considered only the tensor $\eta^S$ to derive the force, we would have
obtained precisely these three terms but the first couple of them
would have had the same coefficient ($a_2$). As we use the complete
tensor $\tau$ we have instead that some coefficients are redundant:
inspecting Eq.\ (\ref{pseudo}), we see that there are two redundant
coefficients among $c_4, c'_4, d_3$, two redundant coefficients among
$a_2, c_3, c'_3, d_2$, and one redundant coefficient among $a_1, d_1$.

After taking into account that $\nabla\cdot{\bm u} = 0$ and
suppressing gradient terms, only remain the coefficients of the part
of $\tau$ that is traceless in the first and second pair of indices.
Gradient terms are longitudinal and the physical force must be
transverse (solenoidal); but, after removing these terms, the force is
still non transverse and must be projected with the nonlocal operator
${\cal P}$ of Eq.\ (\ref{project}). This operation brings back two
supressed gradient terms, namely, $\nabla(\bO\cdot{\bm \o}) = 0$ and
$\nabla[({\bf \Omega \cdot \nabla})({\bf \Omega} \cdot {\bm u})]$, in
addition to producing nonlocal gradient terms.

Finally, we remark that all the terms in Eq.\ (\ref{pseudo}) coming
from odd-$J$ components of $\tau$, that is, the ones with odd powers
of $\bO$ (with coefficients $b_1$, $b_2$, $c_2$, $c'_2$, $c_4$, $c'_4$
and $d_3$), would not be allowed if isotropy were broken by a polar
vector.

\subsection{Dissipation}

The dissipated power is
\begin{eqnarray}
-\int \! d^3x \;{\bm u}\cdot{\bm f} = 
-\int \! d^3x \;u_i \,\p_j\sigma_{ij} = 
\int \! d^3x \;\p_ju_i\, \sigma_{ij} = \nonumber \\ 
\int \! d^3x \;u_{ij} \eta_{ijmn} u_{mn} +
\int \! d^3x \;u_{ij} \chi_{ijmn} \o_{mn} -
\int \! d^3x \;\o_{ij} \xi_{ijmn} u_{mn} -
\int \! d^3x \;\o_{ij} \zeta_{ijmn} \o_{mn}\,,
\label{dissi}
\end{eqnarray}
where we have assumed that the velocity vanishes on the boundary to
remove the surface integrals, that is, we have assumed that there is
no work made by external sources.

As remarked in Ref.\ \cite{us}, $\eta^A$ does not lead to dissipation;
neither does $\zeta^A$. Moreover, if $\chi_{ijmn} = \xi_{mnij}$, the
respective terms cancel in Eq.\ (\ref{dissi}).  All these
nondissipative components of $\tau$ do not properly belong to the {\em
viscosity} tensor, although they give rise to forces with dynamical
effect. On the other hand, since the dissipation cannot be negative,
we can deduce some positivity conditions on the proper coefficients of
the viscosity tensor: $a_1 > 0$, $-d_1 > 0$, etc.

\section{Axisymmetric form of the response function and 
velocity correlations} 

It is useful to study the symmetry constraints on the response
function and velocity correlations.  Here we determine the most
general axisymmetric forms of these quantities in the small-wavenumber
limit (corresponding to large-scale features).  The theory of
axisymmetric tensors has been the subject of previous analyses of
anisotropic turbulence; in particular, it has been treated in papers
by Chandrasekhar \cite{Chandra} and by Arad et al \cite{Arad}. The
former uses the old formalism of invariant theory whereas the latter
uses the theory of group representations. Unfortunately, both consider
only the application to correlation functions in real space, while we
are interested here in correlation functions in Fourier space
(spectral functions). Therefore, the theory of axisymmetric tensors as
is developed in those references must be adapted to Fourier space.
Actually, the spectral two-point velocity correlation function in
rotating turbulence has already been studied by Cambon and Jacquin
\cite{Cambon1} and we shall use their results.

\subsection{Axisymmetric form of rank-two tensors}
\label{AxiT2}

We consider a second-rank tensor that depends on the wave vector ${\bm
k}$ (since we use Fourier space), in addition to the angular velocity
$\bO$.  The general form of such a tensor as a linear combination of
the tensorial products $k_i k_j$, ${\O}_i \O_j$, $k_{i} \O_{j}$,
$\O_{i} k_{j}$ and the unit tensor $\delta_{ij}$ is
\begin{equation}
T_{ij}({\bm k}) = A k_i k_j + B \O_i \O_j + C k_{i} \O_{j} + C' \O_{i} k_{j}
+ E \delta_{ij},
\label{Chand1}
\end{equation}
where $A, B, C, C', E$ are arbitrary functions of $k$, $\O$ and ${\bm
k}\cdot\bO$.  However, a more general expression results upon
introducing the unit antisymmetric tensor or, equivalently, the vector
${\bm n} = {\bm k} \times \bO$ (assuming that ${\bm k}$ and $\bO$ are
not parallel) and the corresponding tensor products:
\begin{eqnarray}
T_{ij}({\bm k}) &=&A k_i k_j  + C k_i \Omega_j + D k_i n_j
\nonumber
\\
&+&
C' \Omega_i k_j  + B \Omega_i \Omega_j + F \Omega_i n_j
\nonumber
\\
&+&
D' n_i k_j  + F' n_i \Omega_j + G n_i n_j
\; .
\label{Chand2}
\end{eqnarray}
This expression with nine coefficients is the most general one,
because any vector (to be included in a tensor product) can be
expressed as a linear combination of ${\bm k}$, $\bO$ and ${\bm n}$.

We remark that expression (\ref{Chand1}) corresponds to the ordinary
quadratic form of Ref.\ \cite{Chandra}, where the terms with the unit
antisymmetric tensor are named ``skew" forms. This name refers to its
reflection (or parity) character: if the two vectors employed in the
tensor products of the ordinary quadratic form are polar, this form is
parity invariant (even parity), whereas the skew forms change sign
under reflections (odd parity) since the vector product is axial.  In
our case, we begin with a polar vector ${\bm k}$ and an {\em axial}
vector $\bO$, so their vector product ${\bm n}$ is polar.  Hence, the
terms of Eq.\ (\ref{Chand2}) that change sign under reflections are
the ones with only one $\bO$.

Instead of the basis formed by ${\bm k}$, $\bO$ and ${\bm n}$,
it may be more convenient to use an orthonormal basis. 
Any couple of linearly independent vectors 
determine an orthonormal basis, in particular, 
the two vectors ${\bm k}$ and ${\bf\O}$ lead to the one given 
by ${\bm k}/k$, ${\bm e}^{(1)} = {\bm k}
\times {\bf\O}/\left|{\bm k} \times {\bf\O}\right|$ and ${\bm e}^{(2)} =
{\bm k} \times {\bm e}^{(1)}/\left|{\bm k} \times {\bm e}^{(1)}\right|$
\cite{Cambon1}. Note that the vector ${\bf\O}$ is axial, 
so ${\bm e}^{(1)}$ is polar but ${\bm e}^{(2)}$ is
axial. Interchanging the role of ${\bf\O}$ and ${\bm k}$, we 
get a different orthonormal basis, with the vector ${\bm e}^{(1)}$ 
in common:
${\bm e}^3 = {\bf\O}/\O$, 
${\bm e}^1 = {\bm k}\times {\bf\O}/\left|{\bm k} \times {\bf\O}\right|
= {\bm e}^{(1)}$ 
and 
${\bm e}^2 = {\bf\O}\times {\bm e}^1/\left|{\bf\O}\times {\bm e}^1\right|$.
This basis (which we denote by superindices without parentheses)
is more adequate. Note that both 
${\bm e}^1$ and ${\bm e}^2$ are polar. 

Any rank-two tensor can be expressed in the latter basis as
\begin{equation}
T_{ij} = \T_{pq} e^p_i e^q_j
\; .
\label{2ten}
\end{equation}
There are three pieces in $T_{ij}$ that are independent under
rotations: the trace, the antisymmetric part and the traceless
symmetric part. This is the Clebsch-Gordan decomposition of the vector
tensor product into the irreducible representations $J=0$, $J=1$ and
$J=2$ of the rotation group.  However, to classify the behaviour of
the components of the second-rank tensor under rotations around
${\bf\O}$, that is, under the two-dimensional rotation subgroup $O(2)$
of the full rotation group $O(3)$, it is best to use the given basis
(components $\T_{pq}$).  The irreducible one-dimensional
representations of $O(2)$ are complex, labeled by an integer $M$ ($-J
\leq M \leq J$). The real irreducible representations are labeled by
$|M|$ and are two-dimensional (except the scalar $M=0$ representation)
\cite{Ham}.  We have that ${\bm e}^3$ is a scalar, and $\{{\bm
e}^1,{\bm e}^2\}$ form the real $|M|=1$ representation.  Consequently,
$\T_{33}$ is the scalar $M=0$ representation,
$\T_{13},\T_{23},\T_{31},\T_{32}$ belong to two $|M|=1$
representations, and the remaining components in the $2 \times 2$
block matrix can be subdivided into its trace ($M=0$), its
antisymmetric part ($M=0$), and its taceless symmetric part ($|M|=2$).
Furthermore, it is not difficult to ascribe each $M$ representation to
a definite $J$ representation.

\subsection{Axisymmetric form of the response function}

The response function is defined by
\begin{equation}
G_{ij}({\bm k},\omega)
= \left.
\frac{\d\langle u_{i}({\bm k},\omega) \rangle}{\d f_{j}({\bm k},\omega)}
\right|_{{\bm f} = {\bf 0}}
\label{resp}
\end{equation}
(introducing a non-random part in the external forcing ${\bm f}$).  So
we can write, at linear order in ${\bm f}$,
\begin{equation}
\langle {u}_{i}({\bm k},\omega) \rangle = 
G_{ij}({\bm k},\omega){f}_{j}({\bm k},\omega)
\; .
\label{resp2}
\end{equation}
Conversely, 
\begin{equation}
G^{-1}_{ij}({\bm k},\omega)
\langle {u}_{j}({\bm k},\omega) \rangle = {f}_{i}({\bm k},\omega),
\end{equation}
which tells us, on account of Eq.\ (\ref{force}), that the quadratic
term in the expansion of $G^{-1}_{ij}({\bm k},0)$ in powers of ${\bm
k}$ is related with the viscosity tensor. To be precise, we have that
\begin{equation}
g_{ijmn} := -\frac{1}{2}
\left.\frac{\p^2 G_{ij}^{-1}}{\p k_m \p k_n}\right|_{{\bm k} = {\bf 0}}
= \frac{1}{2}(\tau_{imjn} + \tau_{injm})\, ,
\label{g-eta}
\end{equation}
defining what we call the (four-rank) response tensor.  This tensor
has symmetry in the pair $mn$, so it has 54 independent components,
while $\tau$ (in the generic case) has 81 components. Indeed, the 27
components of the tensor $\frac{1}{2}(\tau_{imjn} - \tau_{injm})$ do
not contribute to the response function.

\subsubsection{Mapping the viscosity tensor to
the response tensor}

We can take in Eq.\ (\ref{g-eta}) $\eta, \chi, \xi$ or $\zeta$ for
$\tau$.  On the other hand, $g_{ijmn}$ can be decomposed into
$ij$-symmetric and $ij$-antisymmetric parts, corresponding to the
respective parts of the response matrix.  Therefore, $g_{ijmn}$ has $6
\times 6 = 36$ components with symmetry in both pairs (belonging to
the $S$ representation) and $3 \times 6 = 18$ components with
antisymmetry in the first pair and symmetry in the second pair
(belonging to the $AS$ representation).

Let us first analyze the components of $g$ coming from $\eta^S$. 
We note that $g_{ijmn} =  (\eta^S_{imjn} + \eta^S_{injm})/2 = 
(\eta^S_{jnim} + \eta^S_{jmin})/2 = g_{jimn}$. Furthermore, 
this tensor is pair-symmetric:
$g_{mnij} =  (\eta^S_{minj} + \eta^S_{mjni})/2 = 
(\eta^S_{imjn} + \eta^S_{jmin})/2 = (\eta^S_{imjn} + \eta^S_{injm})/2 = 
g_{ijmn}$. 
So the 21 pair-symmetric components of $\eta^S$
(representations $_{15}S$ and $_{6}S$) are
transformed by Eq.\ (\ref{g-eta}) into the 21 pair-symmetric 
components of $g$; in particular, the totally symmetric representation 
$_{15}S$ of $\eta^S$ is left invariant.
Given that we can substitute $\eta^S$ by $\zeta^S$ in the preceding 
equations, we conclude that 
the 6 pair-symmetric components of $\zeta^S$ ($_{6}A$) are
transformed by Eq.\ (\ref{g-eta}) into 6 pair-symmetric components of $g$ 
(linear combinations of $_{15}S$ and $_{6}S$).

We also note that if we take $\chi$ for $\tau$ in Eq.\ (\ref{g-eta})
and symmetrize in $ij$, the tensor $g_{ijmn} = \frac{1}{4}
(\chi_{imjn} + \chi_{injm} + \chi_{jmin} + \chi_{jnim})$ is pair
antisymmetric, owing to the symmetry of $\chi$.  An analogous property
is fulfilled by the tensor $g_{ijmn}$ constructed in the same way from
$\xi$.  So the 15 components of $\chi$ from $_{15}S\!A$ or the 15
components of $\xi$ from $_{15}AS$ are transformed by Eq.\
(\ref{g-eta}) into the 15 pair-antisymmetric components of $g$
(representation $_{15}S'$). On the other hand, it can be proved that
the $\chi$ or $\xi$ belonging to representations $_{3}S\!A$ or
$_{3}AS$, respectively, yield vanishing $g$ (they contribute instead
to $\tau_{imjn} - \tau_{injm}$).

As regards the $ij$-antisymmetric part of $g_{ijmn}$, note that
$g_{ijmn} = (\eta^A_{imjn} + \eta^A_{injm})/2 = (-\eta^A_{jnim} -
\eta^A_{jmin})/2 = -g_{jimn}$.  So the 15 components of $\eta^A$
(representation $_{15}S'$) are transformed by Eq.\ (\ref{g-eta}) into
15 components of $g_{ijmn}$ with antisymmetry in $ij$ (${}_{15}AS$).
Given that we can substitute $\eta^A$ by $\zeta^A$ in the preceding
equations, the remaining 3 components of $\zeta^A$ ($_{3}A$) are
transformed into three components of $g_{ijmn}$ forming the
representation $_{3}AS$.  Finally, we also have the mapping $S\!A \ra
AS$ given by $g_{ijmn} = \frac{1}{4} (\chi_{imjn} + \chi_{injm} -
\chi_{jmin} - \chi_{jnim})$ and a similar mapping $AS \ra AS$.

\subsubsection{Axisymmetric form of the response tensor}

The preceding mapping has been established with full generality,
without considering any particular spatial symmetry. If we take
axisymmetry into account, Eq.\ (\ref{g-eta}) provides the linear
relations between the coefficients in the axisymmetric form of
$g_{ijmn}$ and the coefficients in the axisymmetric forms of $\eta$,
$\chi$, $\xi$ or $\zeta$.  The 54 components of $g_{ijmn}$ belong to
the $S$ and $AS$ representations, therefore, $g_{ijmn}$ has $8 + 4 =
12$ coefficients: the axisymmetric form of the components $g_{ijmn}$
with symmetry in the first pair is like the forms of $\eta$ in
(\ref{nus}) and (\ref{nua}), with other coefficients, say $\a_1,
\ldots, \a_5$ , $\beta_1, \beta_2,\beta_3$; the axisymmetric form of
the components $g_{ijmn}$ with antisymmetry in the first pair is like
the form of $\xi$, with other coefficients, say $\g_1, \ldots, \g_4$.

The above-mentioned coefficients in the response tensor can also be
obtained by expanding in powers of ${\bm k}$ the axisymmetric
expression of $G_{ij}^{-1}$ (\ref{Chand2}).  Considering that the unit
antisymmetric tensor does not appear in the tensors with coefficients
$\a_1, \ldots, \a_5$, $\beta_3$, $\g_3$, the corresponding part of
$G_{ij}^{-1}$ is given just by expression (\ref{Chand1}). Then the
coefficients $\a_1, \ldots, \a_5$, $\beta_3$, $\g_3$ must arise by
expanding $A, B, C, C', E$ in powers of ${\bm k}$ such that the total
expression is of second degree in ${\bm k}$.  In particular, $B = B_1
k^2 + B_2 ({\bm k}\cdot\bO)^2$ and $E = E_1 k^2 + E_2 ({\bm
k}\cdot\bO)^2$, while $A$ is a constant and the coefficient functions
of the terms that are of first degree in ${\bm k}$, namely, $C, C'$,
can only be expanded up to the first order (proportional to ${\bm
k}\cdot\bO$).  Therefore, this expansion just doubles the coefficients
of $\O_i \O_j$ and $\d_{ij}$, producing seven coefficients
altogether. We can divide them between the six coefficients arising
from the symmetric $G_{(ij)}^{-1}$ ($C=C'$) and the one corresponding
to the antisymmetric $G_{[ij]}^{-1}$, namely, $C-C'$.

The part of $G_{ij}^{-1}$ that includes the unit antisymmetric tensor,
with coefficients $D,D',F,F',G$ in Eq.\ (\ref{Chand2}), corresponds to
$\beta_1, \beta_2, \g_1, \g_2$ and $\g_4$.  They can be divided into
$\beta_1, \beta_2$ for the symmetric $G_{(ij)}^{-1}$, and $\g_1, \g_2$
and $\g_4$ for the antisymmetric $G_{[ij]}^{-1}$.
 
\subsubsection{Higher-rank axisymmetric response tensors}

An expansion of $G_{ij}^{-1}$ in powers of ${\bm k}$ to orders higher
than the quadratic order yields response tensors similar to $g_{ijmn}$
in Eq.\ (\ref{g-eta}) but of higher rank. Given that $G_{ij}^{-1}$
must be parity symmetric, only even powers of ${\bm k}$ can appear.
For example, the next higher-rank response tensor $g_{ijmnpq}$ is
symmetric in the last four indices and, therefore, has $9 \times 15 =
135$ components, but this number is reduced by the axisymmetry.  The
following higher-rank response tensors are progressively more complex,
of course.

\subsection{Axisymmetric form of the two-point velocity correlation}

Let us introduce the spectral two-point velocity correlation:
\begin{equation}
\langle u_i({\bm k},\omega)\,u_j({\bm k}',\omega') \rangle =
(2\pi)^{4}{\cal U}_{ij}(\omega,{\bm  k})\, \delta(\omega +
\omega') \, \delta^{3}({\bm k} + {\bm k}').
\end{equation}
We have that ${\cal U}_{ij}(\omega,{\bm k}) = {\cal
U}_{ji}(-\omega,-{\bm k})$. Furthermore, transversality implies that
$k_i{\cal U}_{ij} = k_i{\cal U}_{ji} = 0$.  So, in the isotropic case,
the spectral two-point velocity correlation is given in terms of only
one function:
\begin{equation}
{\cal U}_{ij}(\omega,{\bm  k}) = {\cal P}_{ij}({\bm k}) 
{\cal U}(\omega,{\bm  k}).
\end{equation}
Taking into account that equal-time correlations are more useful, let
us define
\begin{equation}
{\cal U}_{ij}({\bm  k}) = \int \frac{d\omega}{2\pi}\,
{\cal U}_{ij}(\omega,{\bm  k})
\label{Uij}
\end{equation}
(assuming that the integral is convergent), so that
\begin{equation}
\langle u_i({\bm k},t)\,u_j({\bm k}', t) \rangle =
(2\pi)^{3}{\cal U}_{ij}({\bm  k})\, 
\delta^{3}({\bm k} + {\bm k}').
\end{equation}
As demonstrated in section \ref{AxiT2}, the general axisymmetric
rank-two tensor has nine independent coefficient functions, but the
transversality conditions reduce their number.  The number of
independent conditions is five, so just four coefficient functions
remain independent, namely, the ones corresponding to the tensor
products of the transverse vectors ${\bm e}^{(1)}$ and ${\bm
e}^{(2)}$.  It is convenient to use the basis corresponding to
circular polarizations ${\bm N} = {\bm e}^{(1)} - i {\bm e}^{(2)}$,
${\bm N}^* = {\bm e}^{(1)} + i {\bm e}^{(2)}$, so that the resulting
tensor can be written as \cite{Cambon1}:
\begin{equation}\label{Q}
{\cal U}_{ij}({\bm  k}) =  e({\bm  k}) {\cal P}_{ij} + 
\Re[z({\bm  k}){N}_i {N}_j ] 
+ i h({\bm  k}) \epsilon_{ijl}\frac{k_l}{k^2}.
\end{equation}
The quantities $e({\bm k})$ and $h({\bm k})$ are the energy and
helicity spectrum and $z({\bm k})$ is a ``complex deviator''.  They
all are even functions of ${\bm k}$.
The preceding form is equivalent to the form with the 
products ${e}_i^{(1)}{e}_j^{(2)}$, on account that
${N}_i {N}_j = {e}_i^{(1)}{e}_j^{(1)} - 
{e}_i^{(2)}{e}_j^{(2)} - i ({e}_i^{(1)}{e}_j^{(2)} 
+ {e}_i^{(2)}{e}_j^{(1)})$, ${\cal P}_{ij} = 
{e}_i^{(1)}{e}_j^{(1)} + {e}_i^{(2)}{e}_j^{(2)}$, 
and $\epsilon_{ijl}k_l/k = {e}_i^{(1)}{e}_j^{(2)} 
- {e}_i^{(2)}{e}_j^{(1)}$.

Velocity correlations for more than two points lend themselves to be
expressed in similar though more complicated ways.

\section{Connection with some approaches to anisotropic turbulence}
\label{approach}

We have already mentioned that fourth-rank tensors associated to
anisotropic turbulence (but without rotation) have been studied
before; for example, in Ref.\ \cite{DuFri}. More recently, in Ref.\
\cite{jap} has been defined one fourth-rank tensor for a flow with a
constant strain rate.  Another fourth-rank tensor is defined in Ref.\
\cite{Proca} in connection with the linearization of a closure
equation in the presence of weak anisotropy.  We now explore
connections between our anisotropic viscosity tensor and those
tensors.

The fourth-rank tensor $C_{ijmn}({\bm k})$ of Ref.\ \cite{jap}
expresses proportionality between the contribution to the correlation
${\cal U}_{ij}({\bm k})$ (defined in Eq.\ (\ref{Uij})) from anisotropy
and the constant strain rate producing the anisotropy:
\begin{equation}
\d{\cal U}_{ij}({\bm  k}) = C_{ijmn}({\bm k})\, u_{mn}\,, 
\label{japC}
\end{equation}
where the strain rate $u_{mn}$ is constant.  The Reynolds stress
tensor
\begin{equation}
\langle u_i({\bm x},t)\,u_j({\bm x},t) \rangle
= \int \frac{d^3k}{(2\pi)^3}\,
{\cal U}_{ij}({\bm k}) = {\cal U}_{ij}
\label{Rey}
\end{equation}
has a deviatoric part that, according to Eq.\ (\ref{japC}), is
proportional to the strain rate, the proportionality constant being
the integral of the tensor $C_{ijmn}({\bm k})$:
\begin{equation}
\d{\cal U}_{ij} = \int \frac{d^3k}{(2\pi)^3}\, \d{\cal U}_{ij}({\bm k})
= \int \frac{d^3k}{(2\pi)^3} \, C_{ijmn}({\bm k}) \;u_{mn}\,.
\label{dU}
\end{equation}
It is also possible to assume that the Reynolds stress tensor
(\ref{Rey}) and the strain rate have some mild dependence on the
spatial coordinate ${\bm x}$.  The corresponding generalization of
Eq.\ (\ref{dU}) is a phenomenological (mean-field) closure relation
that can be justified with a multi-scale method applied to the
Navier-Stokes equation linearized with respect to the ${\bm
x}$-dependent (large-scale) mean flow \cite{DuFri}.  Comparing this
mean-field relation with Eq.\ (\ref{newton}), we deduce a relation
between our viscosity tensor $\eta_{ijmn}$ and $C_{ijmn}({\bm k})$,
namely,
\begin{equation}
\eta_{ijmn}  = \r \int \frac{d^3k}{(2\pi)^3} \, C_{ijmn}({\bm k}).
\end{equation}
We must note, however, that the form of $C_{ijmn}({\bm k})$ 
in terms of projectors ${\cal P}_{ij}({\bm k})$ proposed 
in Ref.\ \cite{jap} leads to the usual isotropic $\eta_{ijmn}$. 
Indeed, one needs an additional quantity, such as the vector $\bO$, 
to define an anisotropic viscosity.
 
More sophisticated closure schemes involve relations between the three
and two-point velocity correlation functions. The Navier-Stokes
equation leads to an equation involving these two types of
correlations, first derived by von K\'arm\'an and Howarth assuming
isotropy. Chandrasekhar \cite{Chandra} developed a theory of
axisymmetric tensors to generalize this equation to axisymmetric
turbulence. As remarked by Frisch \cite{Frisch}, it is easy to derive
a fully anisotropic version of the K\'arm\'an-Howarth equation, which
he calls the K\'arm\'an-Howarth-Monin equation.  In Ref.\
\cite{Proca}, the Fourier transform of this equation is used as the
basis of a closure scheme related with the direct interaction
approximation (DIA), in which the function ${\cal U}_{ij}({\bm k})$
satisfies (in stationary conditions) a nonlinear equation:
\begin{equation}
D_{mn}({\bm  k}) \equiv I_{mn}({\bm  k})
- \nu k^2 {\cal U}_{mn}({\bm  k}),
\label{Proca-eq}
\end{equation}
where $I_{mn}$ is an integral operator quadratic in ${\cal
U}_{ij}$. In addition, we have introduced an external random forcing,
absent in Ref.\ \cite{Proca}, which is Gaussian {\em and} white in
time and, hence, is represented by the spectral two-point correlation
$D_{mn}({\bm k})$ (see Ref.\ \cite{Frisch} for a general description
of closure equations).  If the molecular viscosity $\nu$ vanishes, the
external forcing is not necessary, as in Ref.\ \cite{Proca}.  However,
the introduction of $D_{mn}$ allows us to substitute the four-rank
tensor defined in Ref.\ \cite{Proca} by a four-rank tensor more useful
to connect with the viscosity tensor.

If the forcing is isotropic, we expect that Eq.\ (\ref{Proca-eq}) has
isotropic solutions.  One can then linearize this equation around an
isotropic solution, namely,
\begin{equation}
\d D_{mn}({\bm q}) = \left.
\frac{\d D_{mn}({\bm q})}{\d{\cal U}_{ij}({\bm  k})}
\right|_{{\cal U}_{ij} =  e {\cal P}_{ij}}
\d{\cal U}_{ij}({\bm  k}), 
\end{equation}
where $\d D_{mn}$ represents an anisotropic perturbation of an
isotropic forcing such that $D_{mn} = D {\cal P}_{mn}$ (note that
isotropy implies that ${\cal U}_{ij} = e\, {\cal P}_{ij}$, according
to Eq.\ (\ref{Q})).  The solution of this linear equation is obtained
by inverting the matrix of pairs of indices, deriving a sort of
tensorial response function,
\begin{equation}
G_{ijmn}({\bm k},{\bm q})
= \left.
\frac{\d{\cal U}_{ij}({\bm  k})}{\d D_{mn}({\bm q})}
\right|_{D_{mn} = D {\cal P}_{mn}},
\label{t-resp}
\end{equation}
which measures the response to the anisotropic perturbation.
Considering the role of the molecular kinematic viscosity $\nu$ in
Eq.\ (\ref{Proca-eq}), we can tentatively define a kinematic viscosity
tensor as
\begin{equation}
\nu_{ijmn}  = -\frac{1}{2}
\left.\frac{\p^2}{\p k_l \p k_l}
\int \! d^3q\; G^{-1}_{ijmn}({\bm k},{\bm q})
\right|_{{\bm k} = {\bf 0}}.
\end{equation}
This relation between a tensorial viscosity and a response tensor is
an alternative to Eq.\ (\ref{g-eta}), valid when we replace the
original Navier-Stokes equation with the closure Eq.\
(\ref{Proca-eq}).  However, the actual computation of $\nu_{ijmn}$
necessarily leads to an isotropic tensor, since there is nothing in
Eq.\ (\ref{t-resp}) capable of breaking rotation invariance.

\section{Conclusions}

We have applied symmetry principles to homogeneous turbulence
subjected to uniform rotation, focusing on the four-rank tensor
defining the linear relation between the stress tensor and the
velocity derivatives, which we call the ``viscosity tensor''. The most
general tensor comprises five parts:
\begin{itemize}
\item a tensor $\eta^S$ symmetric by pairs of indices and pair
symmetric, accounting for the usual proportionality relation between
the (anisotropic) stress and the strain rate;
\item a tensor $\eta^A$ symmetric by pairs and pair antisymmetric
embodying a new relation between the stress and the strain rate,
typical of rotating fluids, since it does not lead to dissipation;
\item a tensor $\chi$ symmetric in the first pair of indices and
antisymmetric in the second, which accounts for a stress tensor
coupling to vorticity;
\item a tensor $\xi$ antisymmetric in the first pair of indices and
symmetric in the second, which accounts for the antisymmetric part of
the stress tensor (angular momentum non-conserving) that couples to
the strain rate.
\item a tensor $\zeta$ antisymmetric in both pairs of indices which
accounts for the antisymmetric part of the stress tensor that couples
to the vorticity.  This tensor can be further decomposed into
pair-symmetric and pair-antisymmetric parts, like $\eta$.
\end{itemize}
Group theory helps us to find the linearly independent components (as
representations of the linear group) in each part.  Every part adopts
a particular axisymmetric form, which can be deduced from the
examination of the decomposition of the linearly independent
components under rotations.

This variety of components of the ``viscosity tensor'' is reflected in
the various effective forces that arise from them. Some of these are
longitudinal, that is, they are the gradient of a potential, and
therefore do not contribute to the transverse rotating fluid
equation. However, they arise from the turbulent state and contribute
to the equation that determines the equilibrium state, so they may
have practical relevance.  Already at first order in $\O$, we have the
potential $\bO\cdot{\bm\o}$, which reminds us of the spin-orbit
coupling of atomic physics. At higher orders in $\O$, we find more
complicated potentials.

Although the most general four-rank ``viscosity tensor'' includes
terms that lead to the stress coupling to vorticity and not conserving
angular momentum, one may wonder if they are really necessary.  If we
take the criterium of having the most general axisymmetric {\em
transverse} force, we could remove redundant coefficients in Eq.\
(\ref{pseudo}): this equation has nine terms but includes the 14
coefficients of the traceless viscosity tensor.  For example, we could
remove all the coefficients belonging to $\zeta$, and the couple $c_3,
c_4$ (belonging to $\chi$) or the couple $c'_3, c'_4$ (belonging to
$\xi$), but not $c_2$ or $c'_2$.  Hence, we conclude that the only
part of the viscosity tensor that can be neglected is $\zeta$, but
$\chi$ and $\xi$ must be present.  So we still have that the stress
couples to vorticity and does not conserve angular momentum (in the
rotating frame).

Finally, in our analysis of the four-rank tensors defined in some
approaches to non-rotating anisotropic turbulence, we have seen there
are similarities with our viscosity tensor in their definition and,
therefore, that the respective definitions can be connected. However,
the lack of specification of a quantity that breaks rotation
invariance precludes that the actual values of the tensors
corresponding to non-rotating turbulence can be anisotropic. It is
possible, nevertheless, to provide such a symmetry-breaking quantity:
for example, an anisotropic noise spectral correlation $D_{mn}$.  In
particular, one can introduce an axisymmetric $D_{mn}$ by postulating
the presence of a global vector (of unknown origin), as in the
perturbative approach of Ref.\ \cite{Slov}. If this symmetry-breaking 
vector were axial instead of polar, the four-rank tensor 
$\nu_{ijmn}$ of section \ref{approach} should have the same form that our 
$\eta_{ijmn}$.

\section*{Acknowledgments}

I thank C. Molina-Par{\'\i}s for conversations.  My
work is supported by a Ram\'on y Cajal contract and by grant
BFM2002-01014, both of the Ministerio de Ciencia y Tecnolog\'{\i}a.

\appendix

\section*{Resolution of the general four-rank tensor by the symmetry of pairs
of indices}

Let us work out first the resolution of the general four-rank tensor
$T_{ijmn}$ into a sum of tensors of definite symmetry type given by
standard Young tableaux.  Young tableaux indicate certain symmetry
operations performed on the indices \cite{Ham}.  We can consider the
general four-rank tensor as a tensorial product of four vectors and,
therefore, write its resolution as the Clebsch-Gordan decomposition
for the linear group $GL(3)$ of the corresponding direct product:
\begin{eqnarray}
\mbox{\begin{picture}(15,15)(0,0)
\multiframe(3,-2)(10.5,0){1}(10,10){$i$}
\end{picture}}
\otimes
\mbox{\begin{picture}(15,15)(0,0)
\multiframe(3,-2)(10.5,0){1}(10,10){$j$}
\end{picture}}
\otimes
\mbox{\begin{picture}(15,15)(0,0)
\multiframe(3,-2)(10.5,0){1}(10,10){$m$}
\end{picture}}
\otimes
\mbox{\begin{picture}(15,15)(0,0)
\multiframe(3,-2)(10.5,0){1}(10,10){$n$}
\end{picture}}
=
\mbox{
\begin{picture}(45,15)(0,0)
\multiframe(3,-2)(10.5,0){4}(10,10){{\small $i$}}{{\small $j$}}%
{{\small $m$}}{{\small $n$}}
\end{picture}}
\oplus
\mbox{
\begin{picture}(35,25)(0,0)
\multiframe(3,3)(10.5,0){3}(10,10){$i$}{$j$}{$m$}
\multiframe(3,-7.5)(10.5,0){1}(10,10){$n$}
\end{picture}}
\oplus
\mbox{
\begin{picture}(35,25)(0,0)
\multiframe(3,3)(10.5,0){3}(10,10){$i$}{$j$}{$n$}
\multiframe(3,-7.5)(10.5,0){1}(10,10){$m$}
\end{picture}}
\oplus
\mbox{
\begin{picture}(35,25)(0,0)
\multiframe(3,3)(10.5,0){3}(10,10){$i$}{$m$}{$n$}
\multiframe(3,-7.5)(10.5,0){1}(10,10){$j$}
\end{picture}}
\oplus
\mbox{
\begin{picture}(25,25)(0,0)
\multiframe(3,3)(10.5,0){2}(10,10){$i$}{$j$}
\multiframe(3,-7.5)(10.5,0){2}(10,10){$m$}{$n$}
\end{picture}}
\oplus
\mbox{
\begin{picture}(25,25)(0,0)
\multiframe(3,3)(10.5,0){2}(10,10){$i$}{$m$}
\multiframe(3,-7.5)(10.5,0){2}(10,10){$j$}{$n$}
\end{picture}}
\oplus
\nonumber\\
\mbox{
\begin{picture}(25,35)(0,0)
\multiframe(3,3)(10.5,0){2}(10,10){$i$}{$j$}
\multiframe(3,-7.5)(10.5,0){1}(10,10){$m$}
\multiframe(3,-18)(10.5,0){1}(10,10){$n$} 
\end{picture}}
\oplus
\mbox{
\begin{picture}(25,35)(0,0)
\multiframe(3,3)(10.5,0){2}(10,10){$i$}{$m$}
\multiframe(3,-7.5)(10.5,0){1}(10,10){$j$}
\multiframe(3,-18)(10.5,0){1}(10,10){$n$} 
\end{picture}}
\oplus
\mbox{
\begin{picture}(25,35)(0,0)
\multiframe(3,3)(10.5,0){2}(10,10){$i$}{$n$}
\multiframe(3,-7.5)(10.5,0){1}(10,10){$j$}
\multiframe(3,-18)(10.5,0){1}(10,10){$m$} 
\end{picture}}
\;.
\label{gen4tensor}
\\\nonumber
\\\nonumber
\end{eqnarray}
The dimensions of the $GL(3)$ representations on the right-hand side
are: 15 for the totally symmetric representation, 15 for the next
mixed symmetry representation, 6 for the following mixed symmetry
representation, and 3 for the last mixed symmetry
representation. Therefore, we have $81 = 15 + 3 \times 15 + 2 \times 6
+ 3 \times 3.$

We intend to show the correspondence of the preceding tensorial
representations with the tensorial representations selected according
to the symmetry relative to pairs of indices. These are constructed as
direct product of representations:
\begin{eqnarray}
\mbox{
\begin{picture}(25,15)(0,0)
\multiframe(3,-2)(10.5,0){2}(10,10){$i$}{$j$}
\end{picture}}
\otimes
\mbox{
\begin{picture}(25,15)(0,0)
\multiframe(3,-2)(10.5,0){2}(10,10){$m$}{$n$}
\end{picture}}
\;, \quad
\mbox{
\begin{picture}(15,25)(0,0)
\multiframe(3,3)(10.5,0){1}(10,10){$i$}
\multiframe(3,-7.5)(10.5,0){1}(10,10){$j$}
\end{picture}}
\otimes
\mbox{
\begin{picture}(25,15)(0,0)
\multiframe(3,-2)(10.5,0){2}(10,10){$m$}{$n$}
\end{picture}}
\;, \quad
\mbox{
\begin{picture}(25,15)(0,0)
\multiframe(3,-2)(10.5,0){2}(10,10){$i$}{$j$}
\end{picture}}
\otimes
\mbox{
\begin{picture}(15,25)(0,0)
\multiframe(3,3)(10.5,0){1}(10,10){$m$}
\multiframe(3,-7.5)(10.5,0){1}(10,10){$n$}
\end{picture}}
\;, \quad
\mbox{
\begin{picture}(15,25)(0,0)
\multiframe(3,3)(10.5,0){1}(10,10){$i$}
\multiframe(3,-7.5)(10.5,0){1}(10,10){$j$}
\end{picture}}
\otimes
\mbox{
\begin{picture}(15,25)(0,0)
\multiframe(3,3)(10.5,0){1}(10,10){$m$}
\multiframe(3,-7.5)(10.5,0){1}(10,10){$n$}
\end{picture}}
\;.
\\\nonumber
\end{eqnarray}
The corresponding dimensions are: 36 for the symmetric-symmetric, 18
for both the antisymmetric-symmetric and the symmetric-antisymmetric,
and 9 for the antisymmetric-antisymmetric.  The symmetric-symmetric
tensor and the antisymmetric-antisymmetric tensor can be further
resolved into pair-symmetric and pair-antisymmetric components.
Moreover, most of the six resulting representatins are still
reducible.  To find the irreducible representations, we will take
advantage of the Clebsch-Gordan decomposition given by Eq.\
(\ref{gen4tensor}), superimposing on it the symmetry relative to pairs
of indices.

To have symmetry in the first pair, say, we may just symmetrize the
general tensor (\ref{gen4tensor}) over indices $ij$. This operation
inmediatly removes the fourth, sixth, eighth and ninth
representations, which are antisymmetric in $ij$. In other words,
given that
\begin{eqnarray}
\mbox{\begin{picture}(15,15)(0,0)
\multiframe(3,-2)(10.5,0){1}(10,10){$i$}
\end{picture}}
\otimes
\mbox{\begin{picture}(15,15)(0,0)
\multiframe(3,-2)(10.5,0){1}(10,10){$j$}
\end{picture}}
=
\mbox{
\begin{picture}(25,15)(0,0)
\multiframe(3,-2)(10.5,0){2}(10,10){$i$}{$j$}
\end{picture}}
\oplus
\mbox{
\begin{picture}(15,25)(0,0)
\multiframe(3,3)(10.5,0){1}(10,10){$i$}
\multiframe(3,-7.5)(10.5,0){1}(10,10){$j$}
\end{picture}}
\;,
\\\nonumber
\end{eqnarray}
we can resolve the general tensor (\ref{gen4tensor}) into
$ij$-symmetric and $ij$-antisymmetric parts. The former is
\begin{equation}
\mbox{
\begin{picture}(25,15)(0,0)
\multiframe(3,-2)(10.5,0){2}(10,10){$i$}{$j$}
\end{picture}}
\otimes
\mbox{\begin{picture}(15,15)(0,0)
\multiframe(3,-2)(10.5,0){1}(10,10){$m$}
\end{picture}}
\otimes
\mbox{\begin{picture}(15,15)(0,0)
\multiframe(3,-2)(10.5,0){1}(10,10){$n$}
\end{picture}}
=
\mbox{
\begin{picture}(45,15)(0,0)
\multiframe(3,-2)(10.5,0){4}(10,10){{\small $i$}}{{\small $j$}}%
{{\small $m$}}{{\small $n$}}
\end{picture}}
\oplus
\left[
\mbox{
\begin{picture}(35,25)(0,0)
\multiframe(3,3)(10.5,0){3}(10,10){$i$}{$j$}{$m$}
\multiframe(3,-7.5)(10.5,0){1}(10,10){$n$}
\end{picture}}
\oplus
\mbox{
\begin{picture}(35,25)(0,0)
\multiframe(3,3)(10.5,0){3}(10,10){$i$}{$j$}{$n$}
\multiframe(3,-7.5)(10.5,0){1}(10,10){$m$}
\end{picture}}
\oplus
\mbox{
\begin{picture}(25,25)(0,0)
\multiframe(3,3)(10.5,0){2}(10,10){$i$}{$j$}
\multiframe(3,-7.5)(10.5,0){2}(10,10){$m$}{$n$}
\end{picture}}
\oplus
\mbox{
\begin{picture}(25,35)(0,0)
\multiframe(3,3)(10.5,0){2}(10,10){$i$}{$j$}
\multiframe(3,-7.5)(10.5,0){1}(10,10){$m$}
\multiframe(3,-18)(10.5,0){1}(10,10){$n$} 
\end{picture}}
\right]_{(ij)}
\;,
\label{symxx}
\end{equation}
where the right-hand side of the latter equation is the result of
symmetrization over indices $ij$ (unnecessary in the totally symmetric
representation).  The $ij$-antisymmetric part is more complicated but
we will not need it.

We can further resolve Eq.\ (\ref{symxx}) by symmetrizing or
antisymmetrizing over the remaining pair of indices:
\begin{eqnarray}
\mbox{
\begin{picture}(25,15)(0,0)
\multiframe(3,-2)(10.5,0){2}(10,10){$i$}{$j$}
\end{picture}}
\otimes
\mbox{
\begin{picture}(25,15)(0,0)
\multiframe(3,-2)(10.5,0){2}(10,10){$m$}{$n$}
\end{picture}}
=
\mbox{
\begin{picture}(45,15)(0,0)
\multiframe(3,-2)(10.5,0){4}(10,10){{\small $i$}}{{\small $j$}}%
{{\small $m$}}{{\small $n$}}
\end{picture}}
\oplus
\left[
\mbox{
\begin{picture}(35,25)(0,0)
\multiframe(3,3)(10.5,0){3}(10,10){$i$}{$j$}{$m$}
\multiframe(3,-7.5)(10.5,0){1}(10,10){$n$}
\end{picture}}
+
\mbox{
\begin{picture}(35,25)(0,0)
\multiframe(3,3)(10.5,0){3}(10,10){$i$}{$j$}{$n$}
\multiframe(3,-7.5)(10.5,0){1}(10,10){$m$}
\end{picture}}
\right]_{(ij),(mn)}
\oplus
\left[
\mbox{
\begin{picture}(25,25)(0,0)
\multiframe(3,3)(10.5,0){2}(10,10){$i$}{$j$}
\multiframe(3,-7.5)(10.5,0){2}(10,10){$m$}{$n$}
\end{picture}}
\right]_{(ij),(mn)}
\;,
\label{symxxyy}
\\\nonumber
\\
\mbox{
\begin{picture}(25,15)(0,0)
\multiframe(3,-2)(10.5,0){2}(10,10){$i$}{$j$}
\end{picture}}
\otimes
\mbox{
\begin{picture}(15,25)(0,0)
\multiframe(3,3)(10.5,0){1}(10,10){$m$}
\multiframe(3,-7.5)(10.5,0){1}(10,10){$n$}
\end{picture}}
=
\left[
\mbox{
\begin{picture}(35,25)(0,0)
\multiframe(3,3)(10.5,0){3}(10,10){$i$}{$j$}{$m$}
\multiframe(3,-7.5)(10.5,0){1}(10,10){$n$}
\end{picture}}
+
\mbox{
\begin{picture}(35,25)(0,0)
\multiframe(3,3)(10.5,0){3}(10,10){$i$}{$j$}{$n$}
\multiframe(3,-7.5)(10.5,0){1}(10,10){$m$}
\end{picture}}
\right]_{(ij),[mn]}
\oplus
\left[
\mbox{
\begin{picture}(25,35)(0,0)
\multiframe(3,3)(10.5,0){2}(10,10){$i$}{$j$}
\multiframe(3,-7.5)(10.5,0){1}(10,10){$m$}
\multiframe(3,-18)(10.5,0){1}(10,10){$n$} 
\end{picture}}
\right]_{(ij),[mn]}
\;.
\label{symxxzz}
\\\nonumber
\\\nonumber
\end{eqnarray}
Let us analyze Eq.\ (\ref{symxxyy}). Using the definitions of the
Young tableaux, we compute
\begin{eqnarray}
\left[
\mbox{
\begin{picture}(35,25)(0,0)
\multiframe(3,3)(10.5,0){3}(10,10){$i$}{$j$}{$m$}
\multiframe(3,-7.5)(10.5,0){1}(10,10){$n$}
\end{picture}}
+
\mbox{
\begin{picture}(35,25)(0,0)
\multiframe(3,3)(10.5,0){3}(10,10){$i$}{$j$}{$n$}
\multiframe(3,-7.5)(10.5,0){1}(10,10){$m$}
\end{picture}}
\right]_{(ij),(mn)}
=
T_{ijmn} + T_{jimn} + T_{ijnm} + T_{jinm}
- T_{mnij} - T_{mnji} - T_{nmij} - T_{nmji}
\,,
\label{sym[xxyy]}
\\\nonumber
\end{eqnarray}
\begin{eqnarray}
\left[
\mbox{
\begin{picture}(25,25)(0,0)
\multiframe(3,3)(10.5,0){2}(10,10){$i$}{$j$}
\multiframe(3,-7.5)(10.5,0){2}(10,10){$m$}{$n$}
\end{picture}}
\right]_{(ij),(mn)}
&=&
T_{ijmn} + T_{jimn} + T_{ijnm} + T_{jinm}
+ T_{mnij} + T_{mnji} + T_{nmij} + T_{nmji} +
\nonumber\\
&&\frac{1}{2}\left[
- T_{mjin} - T_{mijn} - T_{jmin} - T_{imjn}
- T_{mjni} - T_{minj} - T_{jmni} - T_{imnj}\right.\nonumber\\&&\left.
- T_{inmj} - T_{jnmi} - T_{nimj} - T_{njmi}
- T_{injm} - T_{jnim} - T_{nijm} - T_{njim}
\right].
\label{sym(xxyy)}
\\\nonumber
\end{eqnarray}
If we denote the components of the tensor with symmetry by pairs
[given by the left-hand side of Eq.\ (\ref{symxxyy})] as
\begin{equation}
S_{ijmn} = T_{ijmn} + T_{jimn} + T_{ijnm} + T_{jinm}\,,
\end{equation}
we can write Eq.\ (\ref{sym[xxyy]}) as
$$
{}_{15}S'_{ijmn} = S_{ijmn} - S_{mnij}
$$
and Eq.\ (\ref{sym(xxyy)}) as
$$
{}_{6}S_{ijmn} = S_{ijmn} + S_{mnij} - 
\frac{1}{2}\left[S_{imjn} + S_{inmj} + S_{mjin} + S_{njmi}
\right]\,,
$$ where the left subscript indicates the dimension of the irreducible
linear representations.  On the other hand, the totally symmetric
representation reads
\begin{equation}
{}_{15}S_{ijmn} = S_{ijmn} + S_{mnij} + 
S_{jmin} + S_{imjn} + S_{injm} + S_{jnim}
\;.
\end{equation}

Now, let us analyze Eq.\ (\ref{symxxzz}). We compute
\begin{eqnarray}
\left[
\mbox{
\begin{picture}(35,25)(0,0)
\multiframe(3,3)(10.5,0){3}(10,10){$i$}{$j$}{$m$}
\multiframe(3,-7.5)(10.5,0){1}(10,10){$n$}
\end{picture}}
+
\mbox{
\begin{picture}(35,25)(0,0)
\multiframe(3,3)(10.5,0){3}(10,10){$i$}{$j$}{$n$}
\multiframe(3,-7.5)(10.5,0){1}(10,10){$m$}
\end{picture}}
\right]_{(ij),[mn]}
=
&& T_{ijmn} + T_{jimn} - T_{ijnm} - T_{jinm} +
\nonumber\\
&&\frac{1}{2}\left[
- T_{injm} - T_{imnj} - T_{jnim} - T_{jmni}
- T_{mjni} - T_{minj} - T_{njim} - T_{nijm} +\right.\nonumber\\&&\left.
  T_{inmj} + T_{imjn} + T_{jnmi} + T_{jmin}
+ T_{mijn} + T_{mjin} + T_{nimj} + T_{njmi}
\right],
\label{symxxzz15}
\\\nonumber
\end{eqnarray}
\begin{eqnarray}
\left[
\mbox{
\begin{picture}(25,35)(0,0)
\multiframe(3,3)(10.5,0){2}(10,10){$i$}{$j$}
\multiframe(3,-7.5)(10.5,0){1}(10,10){$m$}
\multiframe(3,-18)(10.5,0){1}(10,10){$n$} 
\end{picture}}
\right]_{(ij),[mn]}
=
&& T_{ijmn} + T_{jimn} - T_{ijnm} - T_{jinm} +
\nonumber\\
&&\frac{1}{2}\left[
  T_{mjni} + T_{minj} + T_{njim} + T_{nijm}
- T_{mjin} - T_{mijn} - T_{njmi} - T_{nimj} +\right.\nonumber\\&&\left.
  T_{jmni} + T_{imnj} + T_{jnim} + T_{injm}
- T_{jmin} - T_{imjn} - T_{jnmi} - T_{inmj}
\right].
\label{symxxzz3}
\\\nonumber
\end{eqnarray}
If we denote the components of the tensor with symmetry in the first
pair and antisymmetry in the second [given by the left-hand side of
Eq.\ (\ref{symxxzz})] as
\begin{equation}
S\!A_{ijmn} = T_{ijmn} + T_{jimn} - T_{ijnm} - T_{jinm},
\label{SA}
\end{equation}
we can write Eq.\ (\ref{symxxzz15}) as
$$
{}_{15}S\!A_{ijmn} = S\!A_{ijmn}  + 
\frac{1}{2}\left[S\!A_{imjn} + S\!A_{inmj} + S\!A_{mjin} + S\!A_{njmi}
\right]\,.
$$
and Eq.\ (\ref{symxxzz3}) as
$$
{}_{3}S\!A_{ijmn} = S\!A_{ijmn}  -
\frac{1}{2}\left[S\!A_{imjn} + S\!A_{inmj} + S\!A_{mjin} + S\!A_{njmi}
\right]\,.
$$

In analogy with Eq.\ (\ref{SA}), we define
\begin{equation}
AS_{ijmn} = T_{ijmn} - T_{jimn} + T_{ijnm} - T_{jinm}\;,
\label{AS}
\end{equation}
and we have analogous definitions for the irreducible linear
representations ${}_{15}AS_{ijmn}$ and ${}_{3}AS_{ijmn}$.

We must also define the tensor with antisymmetry by pairs
\begin{equation}
A_{ijmn} = T_{ijmn} - T_{jimn} - T_{ijnm} + T_{jinm}\;, 
\label{A}
\end{equation}
which can be divided into pair-symmetric and pair-antisymmetric
tensors as
\begin{eqnarray}
{}_{6}A_{ijmn} = A_{ijmn} + A_{mnij}\;, \label{A6}\\
{}_{3}A_{ijmn} = A_{ijmn} - A_{mnij}\;. \label{A3}
\end{eqnarray}
Note that ${}_{6}A$ exactly corrsponds to the second square 
Young tableau of Eq.\ (\ref{gen4tensor}). 

Of course, many of the preceding irreducible representations of the
group $GL(3)$ are reducible with respect to its rotation subgroup
$O(3)$, since the metric $\d_{ij}$ allows one to extract lower-rank
representations by contracting indices. For example, the totally
symmetric tensor contains the representations $J = 4,2,0$
\cite{Ham}. The six-dimensional representation given by ${}_{6}S$ or
${}_{6}A$ contain $J = 2,0$. The 15-dimensional representations given
by ${}_{15}S'$ or ${}_{15}S\!A$ contain $J = 3,2,1$. The irreducible
representations of the group $O(3)$ are the symmetric traceless
tensors \cite{Ham}.  Therefore, it must be possible to express each of
the previous representations in terms of four-rank tensors as
symmetric traceless tensors (of lower rank). For example, for $J = 3$,
from ${}_{15}S'$ we obtain $T_{ljn} = \epsilon_{lim} S'_{ijmn} +
\epsilon_{jim} S'_{ilmn} + \epsilon_{nim} S'_{ijml}$; for $J = 3$,
from ${}_{15}S\!A$ we obtain $T_{pij} = \epsilon_{pmn}\, S\!A_{ijmn} +
\epsilon_{imn}\, S\!A_{jpmn} + \epsilon_{jmn}\, S\!A_{pimn}$. Note
that, in both cases, the components of the $J=2$ representations
vanish and do not contribute to the third-rank tensor, while the
remaining $J=1$ representation must be removed from this tensor by
imposing that it be traceless.

\end{document}